\documentclass[prl,twocolumn,showpacs,floatfix,longbibliography]{revtex4-2}
\usepackage{mathrsfs,braket}
\usepackage{amssymb, amsbsy, amsmath, latexsym, dsfont, array, layout, graphicx, mathrsfs, color, ulem, bm}
\usepackage{amsthm}
\usepackage[colorlinks=true,linkcolor=blue,anchorcolor=red,citecolor=blue, urlcolor=blue]{hyperref}

\begin{document}
\title{Anisotropic Anderson localization in higher-dimensional nonreciprocal lattices}
\author{Jinyuan Shang\textsuperscript{1, 2}}
\affiliation{\textsuperscript{1}Beijing National Laboratory for Condensed Matter Physics, Institute of Physics, Chinese Academy of Sciences, Beijing 100190, China}
\affiliation{\textsuperscript{2}School of Physical Sciences, University of Chinese Academy of Sciences, Beijing 100049, China}
\author{Haiping Hu\textsuperscript{1, 2}}\email{hhu@iphy.ac.cn}
\affiliation{\textsuperscript{1}Beijing National Laboratory for Condensed Matter Physics, Institute of Physics, Chinese Academy of Sciences, Beijing 100190, China}
\affiliation{\textsuperscript{2}School of Physical Sciences, University of Chinese Academy of Sciences, Beijing 100049, China}
\begin{abstract}
Nonreciprocity breaks the symmetry between forward and backward propagation, giving rise to a range of peculiar wave phenomena. In this work, we investigate Anderson localization in higher-dimensional nonreciprocal lattices. Focusing on the two-dimensional Hatano–Nelson model, we uncover anisotropic hybrid modes (HMs) that exhibit skin localization along one direction and Anderson localization along the other. We determine the Anderson transition along different directions via the transfer matrix approach and finite-size scaling of Lyapunov exponents. This allows us to map out mobility edges that separate HMs from normal skin modes and Anderson localized modes (ALMs), revealing an ALM–HM–ALM reentrant transition. Our analysis extends to arbitrary dimensions, and we demonstrate the existence of skin–Anderson transitions on the infinite-dimensional nonreciprocal Bethe lattice using the forward-scattering approximation.
\end{abstract}
\maketitle
\textit{Introduction.} Anderson localization is a paradigmatic wave phenomenon in disordered media, where destructive interference halts wave propagation \cite{Anderson1958,at_rev,B_Kramer_1993,Evers2008,at_book}. The spatial dimensionality and symmetries play a crucial role in determining localization behaviors \cite{Abrahams1979,scaling2,scaling3}. In one and two dimensions (1D/2D), eigenstates are exponentially confined by an infinitesimal weak disorder, whereas in 3D, a disorder-driven transition between localized and extended states can occur. In recent years, non-Hermiticity has emerged as a new ingredient in this context \cite{nhat1,nhat2,nhat_nuft,hd_nhat1,nhat3,nhat4,nh_opscaling,jh_schen,lyx_schen,nhat5,nhat6,nhat7,nhat8,me0,me3,me5,me6,me7,me8,me9,taylor,wz_green,impurity1,impurity2,impurity3,impurity4,impurity5,impurity6,impurity7,photonic_exp,wz_jump,hh_jump,hh_jump2}. Its interplay with disorder has led to a range of exotic behaviors such as scale-free localization \cite{impurity1,impurity2,impurity3,impurity4,impurity5,impurity6,impurity7} and universal spreading dynamics \cite{photonic_exp,wz_jump,hh_jump,hh_jump2}. As non-Hermiticity introduces additional length scales, Anderson transitions can even take place in 1D, governed by novel scaling theories \cite{nhat_nuft,nh_opscaling}.

Beyond Anderson localization, non-Hermitian systems exhibit a distinct boundary accumulation of eigenstates known as the non-Hermitian skin effect (NHSE) \cite{nhse1,nhse2,nhse3,nhse4,nhse5,nhse6,exp1,exp2,exp3,exp4,exp5,exp7}. Originating from point gaps in the complex energy spectrum \cite{point1,point2,point3}, the NHSE reflects the intrinsic nonreciprocity of the underlying system. It is the competition between nonreciprocity and disorder that allows Anderson transitions to occur in 1D. However, most existing studies on disordered non-Hermitian systems have focused on 1D, with only a few exceptions \cite{nhat4,nhat5,hd_nhat1} exploring Anderson transitions in higher dimensions. In 1D, spectral and localization properties are well captured by non-Hermitian Thouless relations \cite{hu_lyapunov}. As both Anderson localization and the NHSE are highly sensitive to spatial dimensionality, their interplay in higher-dimensional systems remains largely uncharted. For instance, in higher-dimensional systems, the NHSE exhibits profound complexity and directional anisotropy \cite{point2,hdnb1,hdnb3,hdnb4,hdnb5,hdnb7,hdnb9}. How do these features influence localization behavior? Can the competition between disorder and nonreciprocity lead to novel types of localized modes?

In this work, we study the 2D Hatano–Nelson model with tunable nonreciprocity. At intermediate disorder strength, we identify three distinct types of eigenstates: skin modes localized at the corners, Anderson localized modes (ALMs) confined in the bulk, and a novel class of eigenstates, termed hybrid modes (HMs), which reside at the boundaries and exhibit skin and Anderson localization along orthogonal directions. Within the transfer matrix framework, we establish the criterion for the Anderson transition, and through finite-size scaling of the Lyapunov exponent (LE), determine the mobility edges that separate HMs from both skin modes and ALMs. We find that the system undergoes an ALM–HM–ALM reentrant transition as disorder increases. Finally, through the forward scattering approximation, we demonstrate that skin–Anderson transitions can occur on the infinite-dimensional nonreciprocal Bethe lattice, despite the fact that nonreciprocity typically suppresses Anderson localization.

\textit{2D Hatano-Nelson model.}
We consider the 2D Hatano-Nelson model with nonreciprocity. The Hamiltonian is 
\begin{align}\label{2dhnmodel}
    H&=\sum_{\langle \bm{r},\bm{r'}\rangle}J_{\bm{r},\bm{r'}}c^\dagger_{\bm{r}}c_{\bm{r'}}+\sum_{\bm{r}}\varepsilon_{\bm{r}}c^\dagger_{\bm{r}}c_{\bm{r}},
\end{align} 
where $\bm{r} = (x, y)$ labels sites on a square lattice, and the first summation runs over all nearest-neighbor pairs. The hopping amplitudes are $J_{(x\pm1,y),(x,y)}=Je^{\pm g_x}$, $J_{(x,y\pm1),(x,y)}=Je^{\pm g_y}$. Here, $g_x$ and $g_y$ represent the nonreciprocity along the $x$- and $y$-directions, respectively. The onsite disorder $\epsilon_{\bm{r}}$ is complex, where the real part represents random potential and the imaginary part corresponds to local gain or loss. Unless stated otherwise, we consider a box distribution of the disorder, i.e., both $\text{Re}[\varepsilon_{\bm{r}}]$ and $\text{Im}[\varepsilon_{\bm{r}}]$ are uniformly drawn from $[-W/2,W/2]$ ($W$ is the disorder strength). $J=1$ is set as the energy unit. We impose open boundary conditions and denote the system size as $L_x\times L_y$.

\begin{figure}[!t]
    \centering
    \includegraphics[width=3.35in]{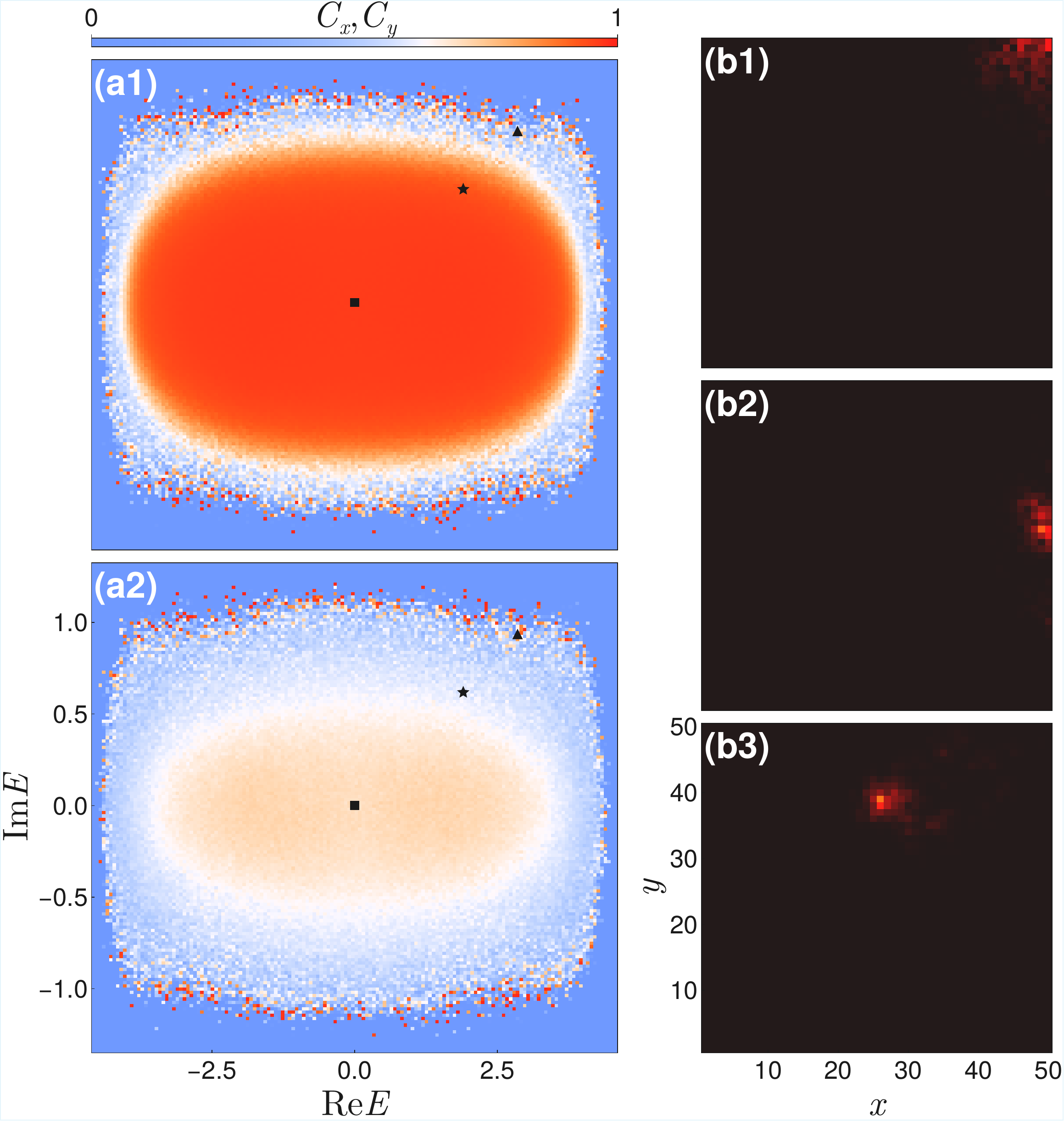}
    \caption{Different types of eigenmodes revealed by the normalized center of mass. (a1) and (a2): Colormaps of the center of mass $C_x$ and $C_y$ along the $x$- and $y$-directions, respectively. 2000 disorder realizations are averaged. (b1)–(b3): Spatial profiles of three representative eigenstates (marked by symbols in (a1) and (a2)) on the square lattice with $L_x=L_y=50$. $g_x=0.3$, $g_y=0.05$, and $W=3$.}
    \label{fig1}
\end{figure}
In the reciprocal limit ($g_x=g_y=0$), the system reduces to a 2D Anderson model with complex disorder. Any finite disorder strength $W$ drives the system into an Anderson insulating phase with all eigenstates becoming ALMs. In contrast, when disorder is absent ($W = 0$), the nonreciprocal model exhibits the NHSE, and eigenstates form skin modes localized at one of the four corners, determined by the signs of $g_x$ and $g_y$. When the two terms coexist, they compete in a way that enables Anderson transitions and gives rise to novel types of eigenmodes. To distinguish between them, we define the normalized center of mass for an eigenstate:
\begin{eqnarray}
C_x=\sum_{x,y} \frac{x |\psi_{x,y}|^2}{L_x};~~C_y=\sum_{x,y} \frac{y |\psi_{x,y}|^2}{L_y}.
\end{eqnarray}
For our analysis with $g_x, g_y > 0$, this metric shows that $C_x$ and $C_y$ are randomly distributed in $[0,1]$ for ALMs, whereas for skin modes at the top-right corner, both $C_x$ and $C_y$ approach unity as the system size increases.

The numerical results from exact diagonalization are shown in Fig. \ref{fig1} The center of mass is obtained by averaging over disorder realizations. In Figs. \ref{fig1}(a1) and (a2), skin modes appear in the central region of the spectrum, where both $C_x$ and $C_y$ are relatively large. Note that exact diagonalization of large non-Hermitian systems is challenging due to numerical errors, leading to noticeable deviations of $C_x$ and $C_y$ from 1. As one moves toward the spectral edges, both $C_x$ and $C_y$ decrease. For each direction, a clear boundary emerges that separates the skin modes from other states. These boundaries generally do not coincide and vary with the nonreciprocity parameters $g_x$ and $g_y$. Figs. \ref{fig1}(b1)–(b3) display representative spatial profiles of three eigenstates. The state from the band center is a corner-localized skin mode. The one near the spectral edge is an ALM in the bulk. The eigenmode in Fig. \ref{fig1}(b2), residing at the right edge, is a HM. It exhibits skin localization along the $x$-direction and Anderson localization along the $y$-direction.

\textit{Anderson transition.}
The transition from skin modes to ALMs can be identified using the transfer matrix method. In 2D, one needs to examine localization along both the $x$- and $y$-directions. Taking the $x$-direction as an example, the strategy is to treat all sites $(x, y)$ with the same $x$ coordinate as a layer, each containing $L_y$ sites along the $y$-axis. The eigenvalue equation can be rewritten as
\begin{align}
    &J_-\psi_{x+1}+J_+\psi_{x-1}+H_x\psi_{x}=E\psi_{x}.
\end{align}
Here $\psi_x$ is a length-$L_y$ vector labeled by layer index $x$, and $H_x$ is the intra-layer Hamiltonian.where $\psi_{x}$ is a vector of length $L_y$, and $H_x$ is the intra-slice Hamiltonian. Specifically, $\left(H_x\right)_{n,m}=\varepsilon_{(x,n)}\delta_{n,m}+Je^{g_y}\delta_{n,m+1}+Je^{-g_y}\delta_{n,m-1}$. The inter-layer couplings $J_+$ and $J_-$ are $L_y\times L_y$ matrices. This eigenvalue problem is then recast into a transfer matrix form:
\begin{eqnarray}
&&  \left(
\begin{array}{c}
  \psi_{x+1} \\
  \psi_x
    \end{array}
    \right)
    =T_x
    \left(
    \begin{array}{c}
        \psi_x \\
        \psi_{x-1}
    \end{array}
    \right),
    \end{eqnarray}
where the $2L_y\times 2L_y$ transfer matrix takes
    \begin{eqnarray}
     T_x
    &&=\left(
    \begin{array}{cc}
        J_-^{-1}(E-H_x) & -J_{-}^{-1}J_+ \\
        1 & 0
    \end{array}
    \right).
\end{eqnarray}
The full transfer matrix $T_L=\prod\limits_{x=1}^{L}T_x$ connects the leftmost and rightmost layers along the $x$ direction. According to Oseledec's ergodic theorem \cite{ergodic}, the eigenvalues of the limiting matrix $\lim\limits_{L\to\infty}\left(T_LT_L^\dagger\right)^{1/2L}$, denoted as $e^{\gamma_{X,i}}$ ($i=1,2,\cdots, 2L_y$), yields $2L_y$ LEs, $\gamma_{X,i}$, along the $x$ direction. 

The LEs are ordered by the magnitude of their moduli for finite $L_y$. Among them, the Anderson transition along the $x$-direction is governed by the essential LE \cite{hu_lyapunov}, denoted $\gamma_X(L_y)$, which explicitly depends on $L_y$. It corresponds to the one closest to zero between $\gamma_{X,L_y}$ and $\gamma_{X,L_y+1}$. In the thermodynamic limit $L_y\rightarrow\infty$, the inverse $1/\gamma_{X}$ gives the localization length of eigenmodes. To perform scaling analysis, we define a rescaled, dimensionless localization length \cite{Slevin2014} $\Lambda_X=\frac{1}{\gamma_X(L_y) L_y}$. The associated beta function is
\begin{eqnarray}
\beta(\ln \Lambda_X)=\frac{d\ln\Lambda_X}{d\ln L_y}.
\end{eqnarray}
Similarly, in the $y$-direction, we construct the corresponding transfer matrix and obtain the essential LE, $\gamma_Y(L_x)$. Information from both directions allows us to determine the skin–Anderson transition of the system.

We now derive the criterion for the transition. For the model (\ref{2dhnmodel}), we apply a similarity transformation: $c^\dagger_{x,y}\to e^{-xg_x-yg_y}c^\dagger_{x,y}$ and $c_{x,y}\to e^{xg_x+yg_y}c_{x,y}$. Under this transformation, the transfer matrix takes $\tilde{T}_x=V_{x+1}T_xV^{-1}_x$, with
\begin{align}
    &V_x=\left(
    \begin{array}{cc}
        e^{xg_x} & 0 \\
        0 & e^{(x-1)g_x}
    \end{array}
    \right)\otimes U,\quad U_{n,m}=e^{n g_y}\delta_{n,m}.
\end{align}
The full transfer matrix becomes $\tilde{T}_{L_x}=V_{L_x+1}T_{L_x} V_1^{-1}=e^{L_xg_x}V_1T_{L_x} V_1^{-1}$. As a result, the LEs shift according to $\tilde{\gamma}_i=\gamma_i-g_x$. Note that this transformation preserves the energy spectrum and eliminates the nonreciprocities, reducing the system to a complex Anderson model with reciprocal hopping.

Suppose we are in the localized regime (of the reciprocal model) and denote its essential LE as $\tilde{\gamma}_X(L_y)$. In the limit $L_y\rightarrow\infty$, the localization length $\tilde{\xi}_{X,\infty}=\lim_{L_y\rightarrow\infty}\frac{1}{\tilde{\gamma}_X(L_y)}$ remains finite. One can always choose $L_y\gg\tilde{\xi}_{X,\infty}$, so that $\tilde{\xi}_X(L_y)\approx \tilde{\xi}_{X,\infty}(1+c/L_y)$ for some constant $c$ and $\tilde{\Lambda}_X^{-1}\approx L_y/\tilde{\xi}_{X,\infty}+a$ with $a$ another constant. We then have:
\begin{align}
\beta=&\frac{d\ln\Lambda_X}{d\ln L_y}=-\frac{d\ln\left(\tilde{\Lambda}_X^{-1}- g_xL_y\right)}{d\ln L_y}\\ \nonumber
\approx&-\frac{1/\tilde{\xi}_{X,\infty}- g_x}{a/L_y+(1/\tilde{\xi}_{X,\infty}- g_x)}.
\end{align}
At the transition point, the beta function vanishes, yielding the transition criterion:
\begin{align}
    g_x=1/\tilde{\xi}_{X,\infty}.
\end{align}
An analogous condition applies along the $y$-direction. We note that a similar criterion has been derived for the 1D Hatano–Nelson model using an imaginary gauge transformation and a wavefunction ansatz \cite{nhat1}. In our 2D case, it is obtained exactly through transfer matrix and renormalization group analysis. This criterion has two key implications. First, when nonreciprocity increases at fixed disorder strength, ALMs are eliminated by the NHSE only when the nonreciprocity exceeds a critical threshold. In other words, for given $g_x$ and $g_y$, the NHSE remains robust against weak disorder. Second, for each spatial direction, there exists a minimum disorder strength below which the NHSE survives. The emergence of HMs originates from anisotropic nonreciprocity along different directions. 
For $d>2$D, HMs are expected to exhibit even richer possibilities. For more general models where nonreciprocity cannot be removed via similarity transformation, numerical methods are required to identify the skin–Anderson transitions in each direction.

\begin{figure}[!t]
    \centering
    \includegraphics[width=3.35in]{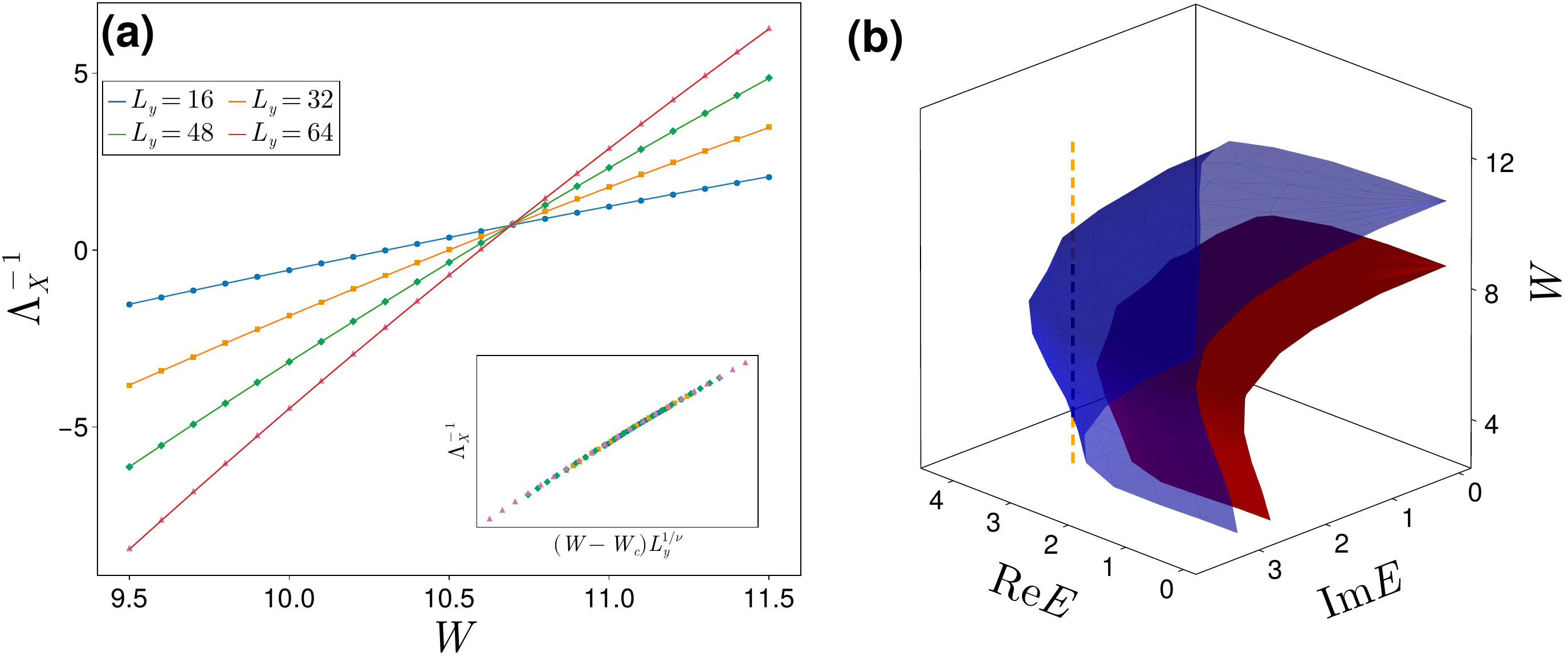}
    \caption{Phase diagram of the 2D Hatano–Nelson model (\ref{2dhnmodel}) with complex disorder. (a) Finite-size scaling of the rescaled Lyapunov exponent (LE), $\Lambda_X^{-1}$, along the $x$-direction, for different vertical lengths $L_y$. The reference energy is set to $E = 0$. Each point is averaged over 100 disorder realizations. The critical point is $W_c \approx 10.69$ with critical exponent $\nu \approx 1$. The fitting parameters are $N_F = 3$ and $N_y = 2$. (b) Mobility-edge surfaces identified from the finite-size analysis (blue/red for the $x/y$ directions, respectively). Model parameters are $g_x = 1$, $g_y = 0.75$.}
    \label{fig:MEs}
\end{figure}

\textit{Finite-size scaling analysis.}
We identify the transition points by performing a finite-size analysis of the rescaled LE. Taking the $x$-direction as an example, we fix the energy and compute the rescaled localization length $\Lambda_x$ as a function of disorder strength for various vertical sizes $L_y$ via the transfer matrix. At the transition, these $\Lambda_x$ curves for different $L_y$ should intersect. We adopt the scaling ansatz $\Lambda_x=F(L_y^{1/\nu}\omega)$, where $F(x)=\sum\limits_{n=0}^{N_F}a_nx^n$ and $\omega=\sum\limits_{n=1}^{N_y}b_n[(W_c-W)/W_c]^n$. Here, $W_c$ and $\nu$ are the critical disorder strength and critical exponent to be determined. We use the standard Levenberg–Marquardt algorithm for the fitting. Figure~\ref{fig:MEs}(a) shows the result at $E=0$. The intersection of curves signals the critical point, and the inset shows the corresponding data collapse. From this analysis, we obtain $W_c\approx10.69$ and $\nu\approx1$.

The finite-size analysis can also be carried out at fixed disorder strength by varying the reference energy and similarly extended to the $y$-direction. This allows us to determine all transition points and mobility edges across both directions and different energies. Numerically, the critical exponent $\nu$ remains close to 1 for all transition points. In Fig.~\ref{fig:MEs}(b), we show the resulting mobility edges for both directions. For clarity, only the first quadrant of the complex energy plane is shown. The full mobility surfaces are symmetric with respect to either the real or imaginary axes. The spectral region between the two separate surfaces corresponds to HMs. The regions that lie outside or inside both mobility surfaces correspond to ALMs and skin modes, respectively.

\begin{figure}[!t]
\centering
\includegraphics[width=3.35in]{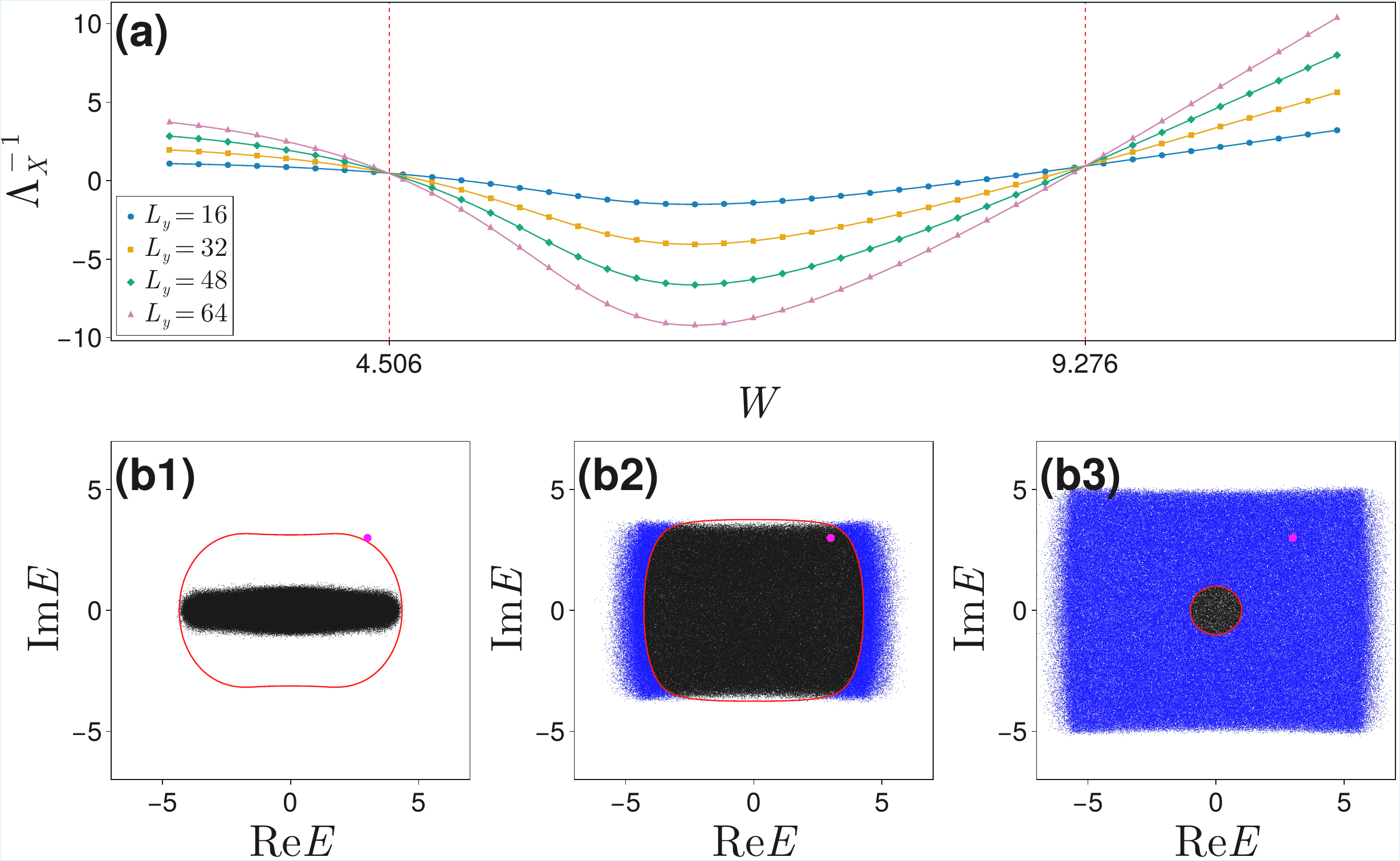}
\caption{Localization reentrance. (a) The rescaled LE, $\Lambda_X^{-1}$ along the $x$ direction versus disorder strength for various vertical sizes $L_y$. The intersection points of these curves yield two critical points at $W_{c_1}\approx4.506$ and $W_{c_2}\approx9.276$. (b1-b3) Energy spectra and mobility edges (red contours) along the $x$ direction in the complex plane for three disorder strengths: $W=3,8,10.64$. The reference energy (marked in magenta) is set as $E=3+3i$.}
\label{fig:dos}
\end{figure}

\textit{Reentrance for intermediate disorder strength.}
If we focus on eigenstates with fixed energy while varying the disorder strength, we find a reentrant localization behavior: ALMs give way to HMs at intermediate disorder, and then return to ALMs as disorder further increases. In Fig. \ref{fig:dos}(a), we track the transition process for a fixed energy $E=3+3i$ by examining the rescaled LE. The two crossing points identify two transitions at $W_{c_1} \approx4.506$ and $W_{c_2} \approx9.276$. The spatial profiles of the corresponding eigenstates indicate that the system host ALMs for $W < W_{c_1}$, HMs for $W_{c_1} < W < W_{c_2}$, and again ALMs for $W > W_{c_2}$. Such reentrance is a consequence of the evolution of mobility edges as disorder increases. In Figs. \ref{fig:dos}(b1)–(b3), we show the positions of the mobility edges in the complex energy plane for the three cases above. Note that for the chosen parameters, Anderson localization remains robust along the $y$-direction.

The reentrance can be understood in terms of spectral density of states. Typically, the skin–Anderson transition first takes place near the spectral edges at weak disorder (yet above the lower threshold discussed earlier). This is because the low density of states near the band edge limits available propagation channels, enhancing interference effects and favoring localization. As disorder increases, the spectrum broadens and the density of states near the edges rises, leading to an outward expansion of the mobility edge, as shown in Fig. \ref{fig:dos}(b2). For a fixed reference energy, an eigenstate that was initially an ALM morphs into a HM. The skin effect persists until disorder becomes dominant. At strong disorder, all eigenstates become Anderson localized, and the spectral density approaches the uniform box distribution in the complex plane.

\textit{Nonreciprocal Bethe lattices.}
Our transfer matrix method and finite-size scaling analysis can be extended to arbitrary finite-dimensional regular lattices to identify the skin–Anderson transition. In $d$D, this requires analyzing the LEs along each spatial direction. Since nonreciprocity generally suppresses Anderson localization, a natural question arises: does the transition still occur in infinite dimensions? Focusing on the Bethe lattice, we provide an affirmative answer using the forward scattering approximation (FWA) \cite{R_Abou-Chacra_1973, PhysRevB.93.054201}.

The Bethe lattice has a tree-like structure where each site connects to $K + 1$ neighbors. For simplicity, we consider uniform disorder with $\epsilon_i \in [-W/2, W/2]$, and nonreciprocal hoppings given by $J_{n\to n\pm1} = -Je^{\pm g}$. Starting from a reference site $a$, the number of sites at distance $n$ is $(K+1)K^{n-1}$. Within the FWA, the Green’s function is expressed as a sum over self-avoiding paths. Owing to the tree geometry, the path $\pi$ from site $a$ to any site $b$ is unique, with length $L(\pi)$. The wave amplitude at site $b$ is then given by
\begin{align}
    \psi(b)=\prod\limits_{i\in\pi}\frac{1}{\epsilon_a-\epsilon_i}(-Je^g)^{L(\pi)}.
\end{align}
Without loss of generality, we set $\epsilon_a=0$. On the Bethe lattice, the number of sites at distance $n$ from site $a$ grows as $K^n$. To analyze localization, we should examine the maximum amplitude decay. Anderson localization occurs if there exists $\alpha > 0$ such that
\begin{align} \label{prob}
\lim\limits_{n\to\infty}\mathbb{P}\left(\max_{b|L(b,a)=n}\frac{\ln\left|\psi(b)\right|^2}{n}<-\alpha\right)=1,
\end{align}
namely, eigenstates decay exponentially with distance, almost surely.

The probability in Eq. (\ref{prob}) can be evaluated by assuming a large deviation form of $\mathbb{P}(\frac{\ln|\psi(b)|^2}{n} > -\alpha) \sim e^{-n\phi(\alpha)}$, where $\phi(a)$ is the rate function, obtainable from its generating function. For large $n$, we have $\mathbb{P}\left(\max_b \frac{\ln |\psi(b)|^2}{n} < -\alpha\right)\sim 1 - \left(1 - e^{-n\phi(\alpha)}\right)^{K^n} \sim 1 - e^{-K^n e^{-n\phi(\alpha)}}$. Starting from the localized regime, the condition for Anderson transition is then given by $\phi(0) = \ln K$. This yields the following estimate for the critical disorder strength:
\begin{align}
    \frac{W_c}{Je^h} = 2eK \ln\left(\frac{W_c}{2Je^h}\right).
\end{align}
In the reciprocal limit ($h \rightarrow 0$), this expression recovers the known result for Hermitian Bethe lattices \cite{PhysRevB.93.054201}. The condition shows that nonreciprocity effectively reduces the influence of disorder and suppresses localization. Although the FWA typically overestimates the transition point, it still captures the existence of Anderson transition in nonreciprocal lattices and qualitatively, predicts that the critical disorder strength increases with nonreciprocity.

\textit{Conclusion and discussion.}
To conclude, we have explored Anderson localization in higher-dimensional nonreciprocal lattices, uncovering a rich interplay among disorder, nonreciprocity, and spatial dimensionality. Focusing on the 2D Hatano–Nelson model, we revealed the emergence of HMs, where skin and Anderson localization coexist along orthogonal directions. Through the transfer matrix method and finite-size scaling of LEs, we identified the transition criterion and mobility edges separating distinct types of eigenstates and demonstrated a reentrant localization transition. In the infinite-dimensional Bethe lattice, the presence of Anderson transitions was confirmed via the FWA.

Our results indicate that Anderson transitions can occur in arbitrary spatial dimensions through the competition between nonreciprocity and disorder. This interplay may give rise to novel localization phenomena in intriguing ways. On the experimental side, the HMs and localization reentrance could be observed in synthetic platforms such as photonic platforms \cite{me0,exp1,photonic_exp} and electrical circuits \cite{impurity3,exp2,exp5,exp7}, where both disorder and nonreciprocal couplings are readily tunable. Future work may explore dynamical aspects of disordered non-Hermitian systems with nonreciprocity, as well as the effects of many-body interactions and nonlinearities on localization behaviors.

\begin{acknowledgments}
This work is supported by the National Key Research and Development Program of China (Grants No. 2023YFA1406704 and No. 2022YFA1405800) and National Natural Science Foundation of China (Grant No. 12474496).
\end{acknowledgments}

\end{document}